\pdfoutput=1
\documentclass[aps,prl,groupedaddress,twocolumn]{revtex4-1}\usepackage[utf8]{inputenc}
\usepackage{graphicx}
\usepackage{braket}
\usepackage{amsmath}
\usepackage{amssymb}
\usepackage[normalem]{ulem}

\usepackage{todonotes}
\presetkeys{todonotes}{inline,backgroundcolor=yellow}{}

\begin{document}

\title{Magic polarization for light shift cancellation in two-photon optical clocks}
\author{Shira Jackson}
\author{Amar C. Vutha}
\email{vutha@physics.utoronto.ca}
\affiliation{Department of Physics, University of Toronto, Toronto, Canada M5S 1A7}
\date{\today}

\begin{abstract}
We find a simple solution to the problem of probe laser light shifts in two-photon optical atomic clocks. We show that there exists a magic polarization at which the light shifts of the two atomic states involved in the clock transition are identical. We calculate the differential polarizability as a function of laser polarization for two-photon optical clocks based on neutral calcium and strontium, estimate the magic polarization angle for these clocks, and determine the extent to which probe laser light shifts can be suppressed. We show that the light shift and the two-photon excitation rate can be independently controlled using the probe laser polarization.

\end{abstract}
\maketitle


Optical clocks have reached an unprecedented level of accuracy, approaching fractional uncertainties of $10^{-18}$ \cite{chou2010,beloy2014,nicholson2015,Grebing2016,huntemann2016,dube2017,Chen2017,mcgrew2018}. These clocks offer powerful tools for precision measurements and tests of fundamental physics \cite{Ludlow2015,Mann2018}, such as searches for dark matter and dark energy \cite{derevianko}, or low-frequency gravitational wave detection \cite{Vutha2015,Kolkowitz2016,Norcia2017}. Optical clocks can be used for geodetic surveys of the earth's gravitational potential \cite{mehlstaubler,Grotti2018,mcgrew2018}. Atomic clocks are also an essential component of the new SI system of units, where the measurement of almost every physical quantity is ultimately related to the accurate measurement of frequency \cite{milton2014,Keller2016}. It is widely expected that the SI second will soon be redefined in terms of optical atomic clocks \cite{McGrew2019}. 

Careful control of systematic errors is necessary to achieve such high levels of accuracy. An important and generic systematic in optical clocks is the light shift -- a shift in the resonance frequency of the atoms due to off-resonant light. For example, one source of light shifts in optical lattice atomic clocks is the lattice laser which must remain on during the measurement. These shifts are usually controlled by choosing a ``magic wavelength" for which the light shifts of the clock ground and excited states are equal \cite{katori2003}. This ensures that the relative energy difference between the two levels is unaffected by the optical lattice, therefore making the clock less sensitive to fluctuations in laser intensity. 

In addition to light shifts from trapping lasers, many optical clocks suffer from light shifts due to the probe laser itself to some extent, because they rely on forbidden atomic transitions. This necessitates careful stabilization of the probe laser intensity, or sophisticated measurement protocols such as hyper-Ramsey schemes \cite{yudin2010,hobson2016,huntemann2016,zanon-willette2017}, autobalanced Ramsey spectroscopy \cite{sanner2018}, or displaced frequency-jump Ramsey spectroscopy \cite{shuker2018} in order to suppress the probe laser light shift. 
Here we consider two-photon optical clocks \cite{Hall1989}. Two-photon clocks using rubidium \cite{Zhu1997,Poulin2002,ducos2002,Edwards2005,Perrella2013,Newman2018,Martin2018}, xenon \cite{Rolston1991,Sterr1995} and silver \cite{Badr2004,Badr2006} have been successfully operated, and clocks using calcium \cite{Hall1989,Vutha2015}, and strontium \cite{Hall1989,hummon} have been proposed. Two-photon optical transitions can be driven using two counter-propagating laser beams at the same frequency, which makes the interrogation of the clock transition insensitive to first-order Doppler shifts and photon recoil shifts \cite{Hall1989}. This offers an attractive path to a compact, field-portable optical clock that does not require strong cooling and confinement in a lattice, and therefore has a simplified architecture. However, the systematic frequency shifts in such clocks are dominated by the light shift due to the probe laser, because of the relatively large laser intensities that are needed to drive the transition (cf. \cite{Martin2018IEEE,Gerginov2018,Newman2018,Martin2018}). Here we show that light shifts due to the probe laser in two-photon optical clocks can be nulled using a specific choice of the laser polarization, removing one of the main obstacles to achieving high accuracy with these clocks. 

The term ``magic polarization" has been used in the context of reducing line broadening and increasing coherence times of hyperfine transitions in alkali atoms \cite{kim2013}, and improving the efficiency of Doppler cooling of a trapped gas \cite{chalopin2018}. It has also been shown that the degree of circular polarization can be used to mitigate the effects of differential hyperpolarizability in optical lattice clocks \cite{katori2015}. Here we use ``magic polarization" to refer to a particular polarization angle of the \emph{probe laser} light for which the differential dynamic polarizability of the two clock states is zero. We find this magic polarization for calcium and strontium two-photon optical clocks by calculating the dynamic polarizabilities of the clock states. We show that the probe laser light shifts can be strongly suppressed, and the fractional frequency uncertainty due to the light shift reduced below $10^{-18}$, using a robust method that is simple to implement. 

The light shift of an atomic energy level due to the electric dipole ($E1$) interaction is $\Delta E_k = -\frac{1}{2} \alpha_k(\omega) \mathcal{E}^2$, where $\alpha_k$ is the dynamic polarizability of the atomic state $\ket{k}$, $\mathcal{E}$ is the electric field amplitude and $\omega$ is the frequency of the electric field. The polarizability $\alpha_k(\omega)$ depends on both the frequency and the polarization of the electric field. The polarizability of a state $k$ can be calculated to leading order in perturbation theory as
\begin{equation}
    \alpha_k(\omega) = \sum_j\frac{|\braket{j|D|k}|^2}{E_j - E_k - \hbar\omega} + \sum_j\frac{|\braket{j|D|k}|^2}{E_j - E_k + \hbar\omega}
    \label{eq:polarizability}
\end{equation}
where $D$ is the $E1$ interaction operator $\vec{d} \cdot \vec{\mathcal{E}}$ and $E_k, E_j$ represent the unperturbed energies of states $\ket{k},\ket{j}$ respectively. The electric dipole operator $\vec{d}$ is a linear combination of three rank-1 spherical tensor operators $d_q$ ($q=0,\pm1$), with relative coefficients depending on the laser polarization.

In all the calculations, the quantization axis was assumed to be defined by a small magnetic field in the $\hat{z}$ direction, and the probe laser was considered to propagate along the $\hat{y}$ direction. We write the electric field as $\vec{\mathcal{E}} = \mathcal{E}_0 \left( \cos{\theta} \, \hat{z} + e^{i\phi} \sin{\theta} \, \hat{x} \right)$, which defines the polarization angles $\theta, \phi$.
The $E1$ operator is therefore $D = \mathcal{E}_0 \left( \cos{\theta} d_0 - \frac{1}{\sqrt{2}} e^{i\phi} \sin{\theta} d_{1} + \frac{1}{\sqrt{2}}e^{i\phi} \sin{\theta} d_{-1} \right)$. 
The light shift for the clock transition is $\delta \omega_\mathrm{LS} = -\frac{1}{2} \; \Delta \alpha_{eg} \; \mathcal{E}_0^2/\hbar$, where $\Delta \alpha_{eg}(\omega_0) = \alpha_e(\omega_0) - \alpha_g(\omega_0)$ is the differential polarizability between the ground and excited clock states, and $\omega_0$ is the frequency of the probe laser.

The matrix elements in Eq.\ (\ref{eq:polarizability}) can be calculated from the reduced matrix elements $\braket{J_k||D||J_i}$, which can in turn be related to experimentally determined oscillator strengths $f_{ik}$ using the relation \cite{drake_2006} $$gf = (2J_i +1) f_{ik} = \frac{2}{3} \frac{m_e \omega_{ik}}{\hbar} \frac{\braket{i||D||k}^2}{e^2},$$ where $gf$ is the symmetric oscillator strength, $f_{ik}$ is the non-symmetric oscillator strength, and $\omega_{ik} = (E_k - E_i)/\hbar$. 

 
Energy levels and oscillator strengths for transitions relevant to the calculation of dynamic polarizabilities are available both from experimental data and \emph{ab initio} theoretical calculations. In calculating the clock state polarizabilities for rubidium, calcium, and strontium two-photon optical clocks, energies and oscillator strengths from the NIST Atomic Spectra Database \cite{NIST2018} and other published sources \cite{NIST2010,guo2010,Vaeck1988,hansen1999,ruczkowski2016} were used.


The energy level differences in these atoms are known to better than 10$^{-5}$, and therefore do not contribute significantly to the uncertainty in the polarizability. The uncertainty is instead dominated by the oscillator strengths. The sum in Eq.\ (\ref{eq:polarizability}) is usually well approximated by only a handful of terms that represent atomic levels $\ket{i}$ which couple strongly to the state $\ket{k}$ and are close in energy. We found that including more than $\sim$10 leading terms to the sum in Eq.\ (\ref{eq:polarizability}) only changed the polarizability by $\sim$ 0.1\%. To estimate the accuracy of the oscillator strengths used as inputs for our magic polarization calculations, we compared the polarizabilities of some reference states against previously published results, as summarized in Table \ref{table:polarizabilities}. For all of these cases, our calculations reproduced the published values reasonably well, indicating that the oscillator strengths of the important transitions are accurate to better than 10\%.


\begin{table}[h]
    \centering
    \begin{tabular}{cccc}
    \hline \hline
    $\lambda$ [nm] & Level & \multicolumn{2}{c}{Polarizability [a.u.]}\\
    \; & & This work & Other results\\
    \hline
     $\infty$ & Ca $4s^2 \; ^1S_0$ & 154.7 & 163.0$^a$ 168.7(16.9) $^b$\\
    $\infty$ & Sr $5s^2 \; ^1S_0$ & 201.6 & 192.5$^c$, 197.2$^d$ \\
    813.4 & Sr $5s^2 \; ^1S_0$ & 295.6 & 278.1$^c$, 286.0$^d$\\
    \hline
    \end{tabular}
    \caption{Calculated polarizabilities (in atomic units), for selected levels in calcium and strontium, that are used to benchmark the input data for the light shift and magic polarization calculations. All calculations are for $m_J$ = 0 sublevels. Our calculations agree with published values to better than 10\% ($^a$Ref.\ \cite{merawa2001}, $^b$Ref.\ \cite{Miller1976}, $^c$Ref.\ \cite{guo2010}, $^d$Ref.\ \cite{JunYe2008}).
    }
    \label{table:polarizabilities}
\end{table}



\begin{figure}
    \centering
    \includegraphics[width=0.8\columnwidth]{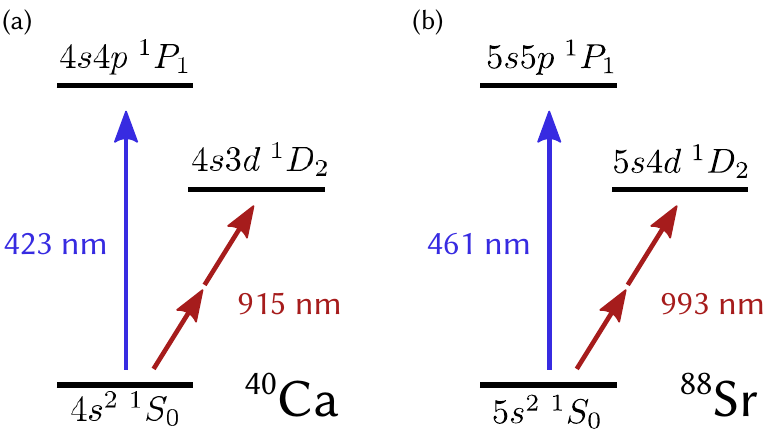}
    \caption{Level structures of (a) $~^{40}$Ca and (b) $~^{88}$Sr. The main cooling line for calcium (strontium) atoms is the ${}^{1}{S}_{0}-{}^{1}{P}_{1}$ cycling transition at 423 nm (461 nm), and the ${}^{1}{S}_{0}-{}^{1}{D}_{2}$ clock transition involves two photons at 915 nm (993 nm).
}
    \label{fig:levels}
\end{figure}

The atomic level structures relevant to two-photon optical clocks in neutral calcium and strontium are shown in Fig.\ \ref{fig:levels}. The strong dipole-allowed ${}^{1}{S}_{0}-{}^{1}{P}_{1}$ cycling transitions are convenient for laser cooling the atoms, in order to increase the interrogation time and reduce the second-order Doppler shift \cite{Hall1989,Vutha2015}. The two-photon clock states are the ground ${}^1S_0$ state and the excited ${}^1D_2, m_J=0$ state. The transition between these states is insensitive to the first-order Zeeman shift. The ${}^{1}{S}_{0}-{}^{1}{D}_{2}$ two-photon clock transition is probed using two laser beams with identical frequencies, but oppositely directed wavevectors. The ground and excited clock states are hereafter denoted as $\ket{g}$ and $\ket{e}$.  

\begin{figure}
    \centering
    \includegraphics[width=\columnwidth]{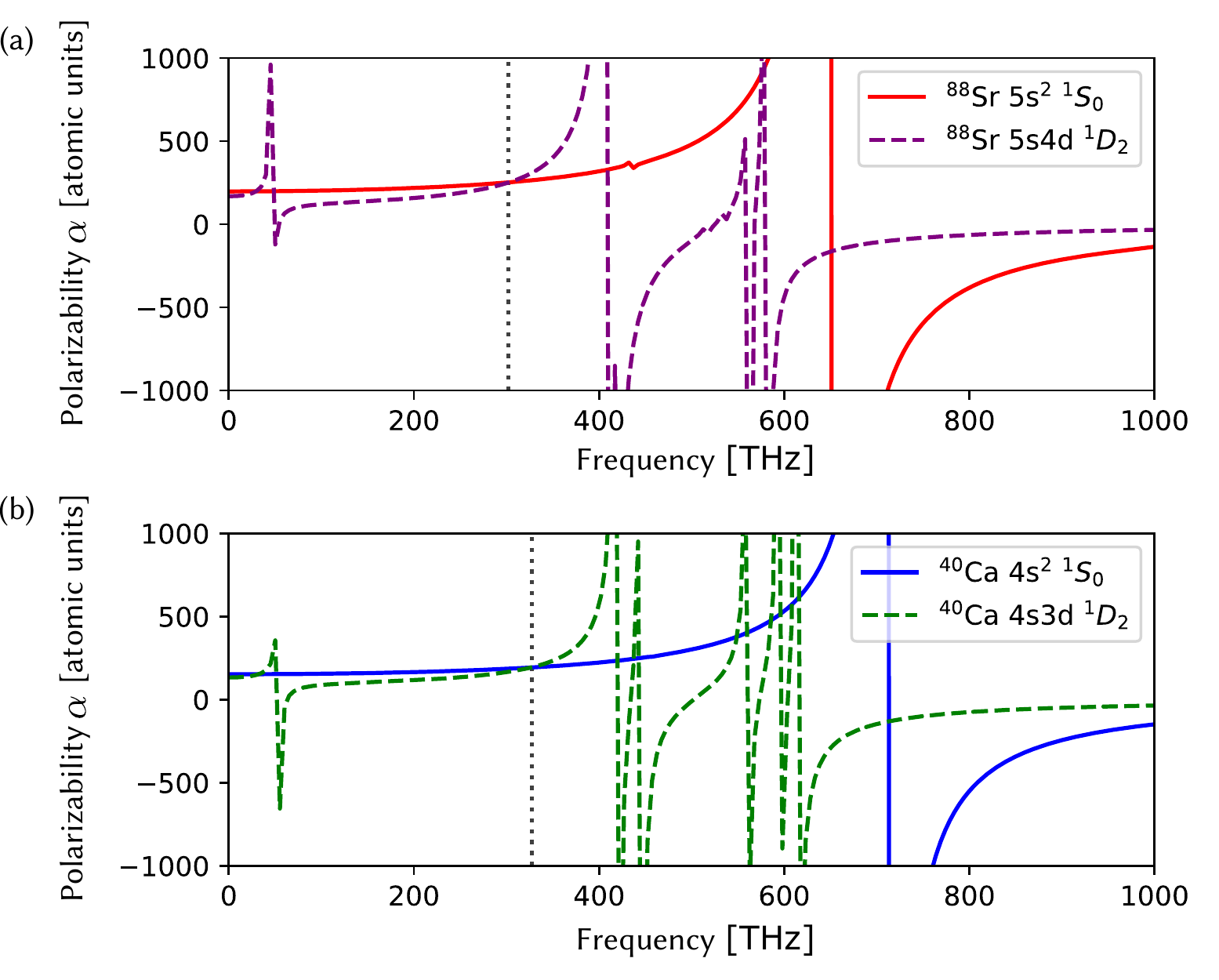}

    \caption{Dynamic polarizabilities of (a) strontium and (b) calcium two-photon clock states at the magic polarization angle $\theta_0$. The vertical dotted grey line indicates the frequency of the clock transition, at which the polarizabilities of the ground and excited states can be made equal.
}
    \label{fig:dyn_pol}
\end{figure}

\begin{figure}
    \centering
    \includegraphics[width=\columnwidth]{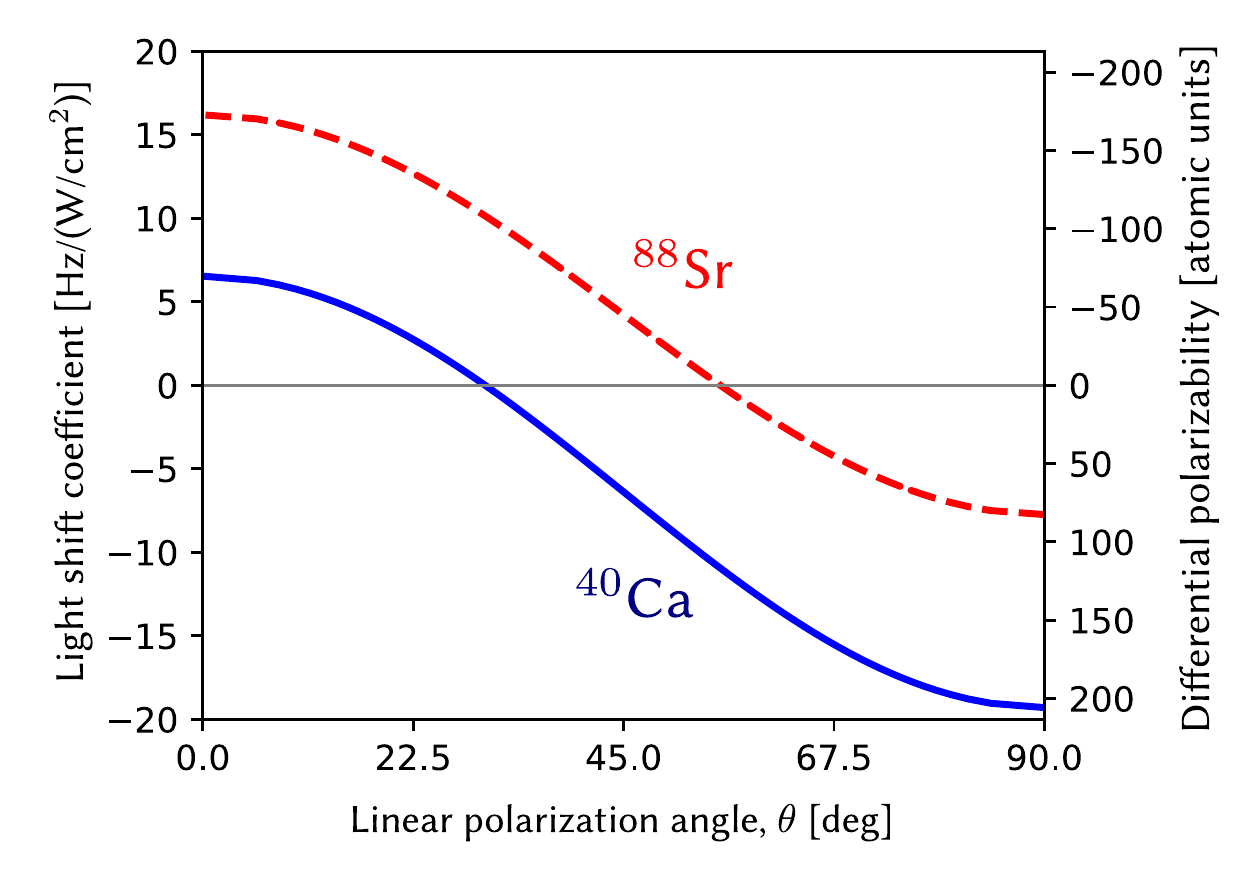}
    \caption{Light shifts for calcium and strontium due to the probe laser at two-photon resonance, as a function of linear polarization angle. $\theta=0^\circ$ denotes z-polarized light, $\theta=90^\circ$ denotes x-polarized light.}
    \label{fig:light_shifts}
\end{figure}

The dynamic polarizabilities calculated for calcium and strontium, in atomic units, as a function of $\omega$ are shown in Fig.\ \ref{fig:dyn_pol}. (In atomic units, a polarizability of 1 is equal to $4 \pi \epsilon_0 a_0^3$, where $\epsilon_0$ is the permittivity of free space and $a_0$ is the Bohr radius.) At the clock transition frequency, the differential polarizability between $\ket{g}$ and $\ket{e}$ (which is proportional to the light shift of the clock transition) is shown as a function of probe laser linear polarization angle in Fig.\ \ref{fig:light_shifts}. Both calcium and strontium two-photon clocks possess a magic polarization at which the differential polarizability is zero. This feature is generic, and we verified that the existence of a magic polarization angle is not affected by $\pm 10\%$ changes in a number of oscillator strength values. 

The existence of a magic polarization in calcium and strontium is related to the light shift of $\ket{e}$. (The polarizability of $\ket{g}$ is independent of the polarization angle, and has positive sign at the two-photon clock transition frequency, which decreases the energy of the ground state for increasing laser intensity.) 
The two odd-parity states closest in energy to $\ket{e}$ are the $4s4p \ (5s5p) \ {}^1P_1^o$ and $3d4p \ (4d5p) \ {}^1D_2^o$ states in calcium (strontium). For $x$-polarization, the probe laser is red-detuned on the ${}^1D_2, m_J=0  \leftrightarrow {}^1D_2^o, m_J=\pm 1$ transition, resulting in a large \textit{positive} polarizability for $\ket{e}$ which exceeds the polarizability of $\ket{g}$. For $z$-polarization,the ${}^1D_2, m_J=0 \leftrightarrow {}^1D_2^o, m_J=0$ transition does not contribute to the polarizability due to angular momentum conservation. The light shift for $z$-polarization is therefore determined by the $4s4p \ (5s5p) \ {}^1P_1^o$ state on which the probe laser is blue-detuned which results in a \textit{negative} polarizability for $\ket{e}$. Between pure $z$- and pure $x$-polarization therefore, there is generically a magic polarization angle $\theta_0$ at which the ground and excited clock states have identical light shifts. Therefore the exact value of the magic angle may be affected by uncertainties in our calculations as described above, or higher-order $\mathcal{O}(\mathcal{E}^4)$ corrections to the polarizabilities, but the existence of a magic angle is a robust result. 

(In passing, we note that we also performed calculations for rubidium two-photon clocks operating on the 778 nm $5S_{1/2} - 5D_{3/2,5/2}$ clock transitions, but did not find a magic polarization angle for degenerate two-photon excitation. This is related to the fact that the two-photon clock frequency happens to be very close to the strongly allowed $5S_{1/2} - 5P_{3/2}$ transition at 780 nm. Therefore the polarizability of the ground state is so large that it entirely dominates the differential polarizability for all polarizations of the probe laser.)

\begin{figure}
    \centering
    \includegraphics[width=\columnwidth]{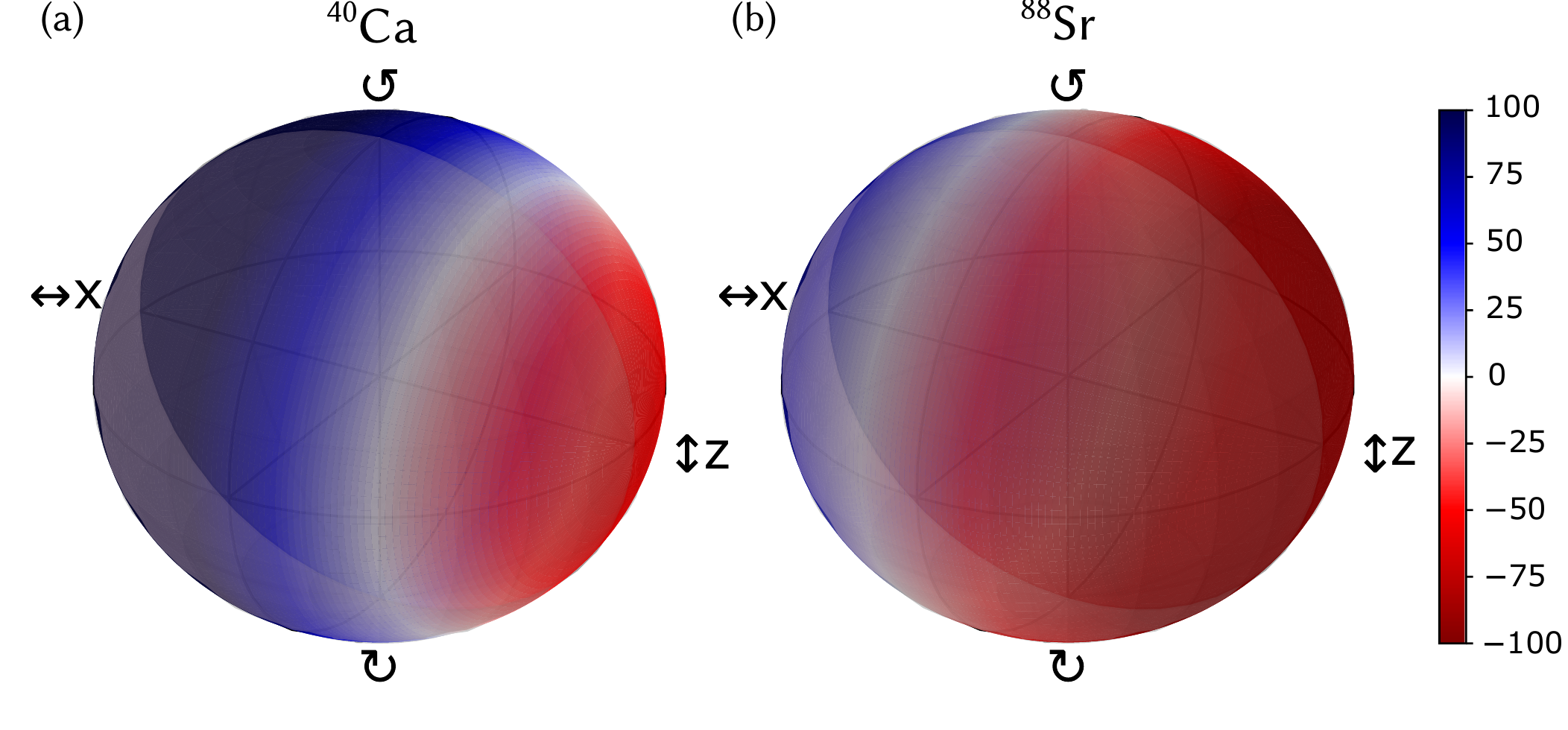}
    \caption{Differential polarizability between the clock states in atomic units for (a) calcium and (b) strontium, plotted on the Poincar\'e sphere. On the sphere, the polarizability is cylindrically symmetric around the axis connecting $x$-polarization to $z$-polarization. The magic polarization angle corresponds to the band with zero differential polarizability.
    }
    \label{fig:poincare}
\end{figure}

For the case considered here, where the clock laser propagates perpendicular to the quantization axis, the differential polarizability does not depend on the ellipticity of the probe laser polarization. As a result the magic polarization angle is robust against any phase differences between the $\hat{x}$ and $\hat{z}$ components of the laser polarization, such as might be accrued in propagation through birefringent optical components (e.g., stressed vacuum viewports). This fact is evident in Fig.\  \ref{fig:poincare}, where the differential polarizability plotted on the Poincar\'e sphere is seen to be independent of $\phi$. 

\begin{figure}
    \centering
    \includegraphics[width=\columnwidth]{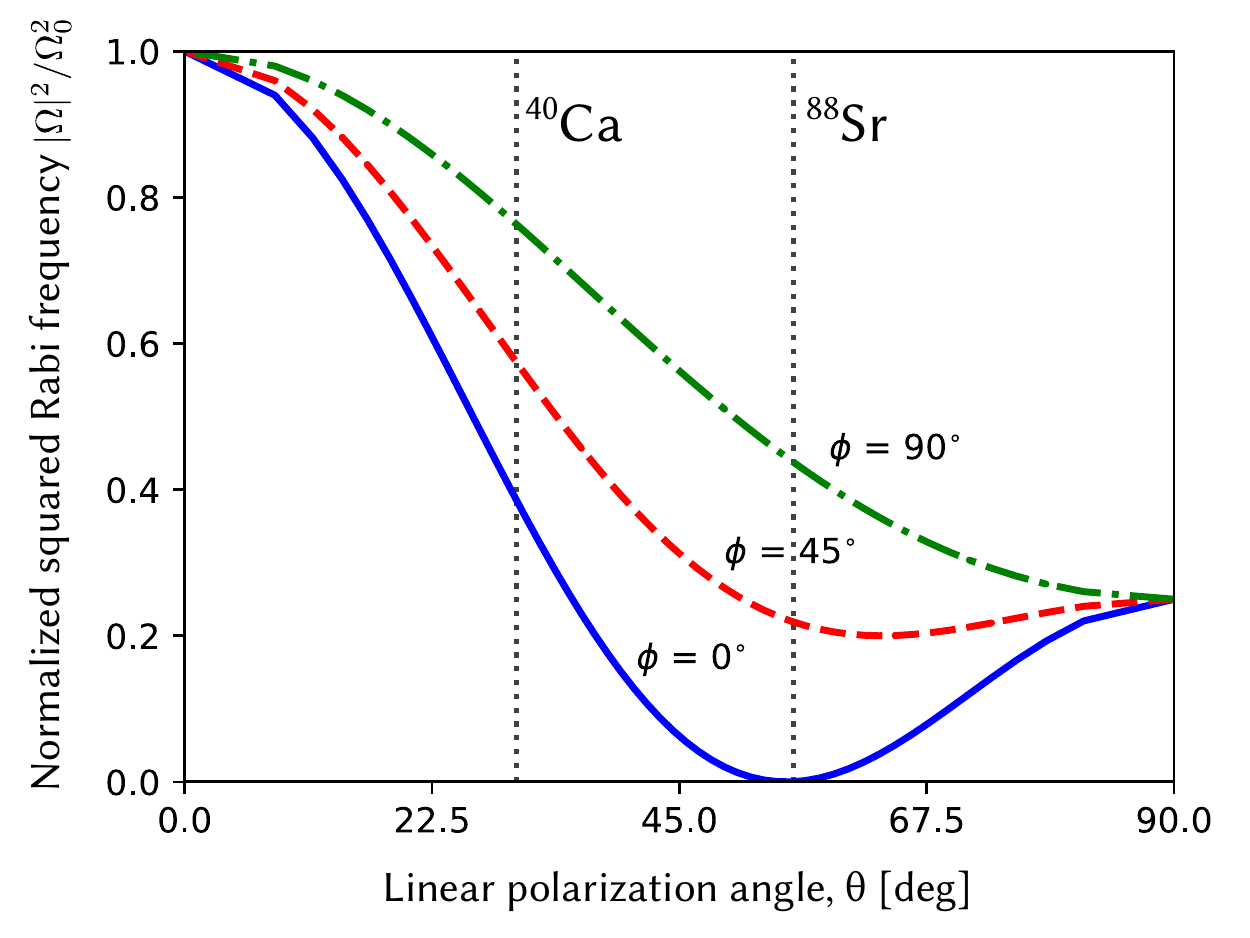}
    \caption{Two-photon Rabi frequencies for strontium and calcium clock transitions as a function of probe laser polarization angle. Rabi frequencies are shown for three values of the angle $\phi$, the phase difference between $\hat{x}$ and $\hat{z}$ components which describes the degree of circular polarization. The plot shows $|\Omega|^2$ normalized to its value at $\theta=0$. The value of $\Omega_0$ = 4.85 (22.1) Hz/(W/cm$^2$) for calcium (strontium). The magic polarizations for calcium and strontium are indicated with a dashed vertical line.}
    \label{fig:rabi}
\end{figure}

For both calcium and strontium the light shift in the vicinity of the magic polarization angle, $\theta_0$, is $\delta \omega_\mathrm{LS} \approx \kappa (\theta - \theta_0)$, with the shift coefficient $\kappa \approx$ -30 Hz/(W/cm$^2$)/rad. From Fig.\ \ref{fig:rabi}, typical probe laser intensities for the calcium and strontium two-photon transitions are expected to be $\sim$ 1 W/cm$^2$. With its polarization set to within 1 mrad of $\theta_0$ over the atomic ensemble (easily accomplished with standard optical components), it is sufficient to stabilize the probe laser intensity to just 1\% to ensure that the systematic error due to the light shift is below 10$^{-18}$. Probing the transition with a magic polarized laser effectively suppresses the light shift by over 2 orders of magnitude, and eliminates one of the main sources of inaccuracy in two-photon optical clocks based on alkaline-earth atoms.

The choice of laser polarization also affects the two-photon Rabi frequency $\Omega$ and the excitation rate $\propto |\Omega|^2$. The two-photon Rabi frequency is $\hbar \Omega = \sum_k \frac{\braket{e|D|k}\braket{k|D|g}}{\hbar \omega_0 - E_k}$. The value of $\Omega$ was numerically calculated for the calcium and strontium two-photon transitions using the same set of oscillator strengths used for the polarizability calculations. The calculated dependence of the Rabi frequency on the probe laser polarization is shown in Fig.\ \ref{fig:rabi}. An analytical expression that accurately predicts the $\theta$ and $\phi$ dependence of the Rabi frequency is derived in the appendix. Intuitively, the dependence of $\Omega$ on laser polarization can be understood as follows: the amplitude for the two-photon transition from $^1S_0, m_J=0 \to ~^1D_2, m_J=0$ is a sum of three interfering amplitudes, for excitation via intermediate $m_J'=0,\pm1$ states. The relative amplitudes and phases of these three terms are set by $\theta$ and $\phi$.  At $\theta = \arccos\left(\frac{1}{\sqrt{3}}\right), \phi=0$ there is complete destructive interference between these three excitation pathways and the Rabi frequency goes to zero (see Section A1 in the Appendix). However, adding a phase difference $\phi$ between the $\hat{x}$ and $\hat{z}$ components of the electric field, destroys this perfect cancellation between amplitudes and prevents the Rabi frequency from becoming zero for any value of $\theta$. As this phase shift does not affect the differential polarizability between the clock states (see Fig.\ \ref{fig:poincare}), the Rabi frequency for the transition can be tuned independently from the magic polarization and optimized as required. For example, controlling $\phi$ allows the strontium two-photon transition to have a large Rabi frequency despite the closeness of its magic polarization angle to the interference minimum at $\approx$ 55$^\circ$, as shown in Fig.\ \ref{fig:rabi}.

In summary, we have described a magic polarization scheme for radically suppressing probe laser light shifts in two-photon optical clocks. Setting the probe laser to a magic polarization angle is sufficient to eliminate systematic errors due to light shifts in calcium and strontium two-photon clocks, opening up a path to compact and portable optical clocks based on simple two-photon transitions. Controlling polarizations to cancel light shifts may also be a useful method for other precision measurements on highly forbidden atomic and molecular transitions.

\emph{Acknowledgments.} This work is supported by the Branco Weiss Fellowship, NSERC, and Canada Research Chairs. S.J. acknowledges support from an Ontario Graduate Scholarship. We are grateful to Eric Hessels and Matthew Hummon for stimulating discussions.

\bibliography{magic_polarization}

\onecolumngrid


\section{Appendix A: Polarization dependence of the two-photon Rabi frequency}
The two-photon Rabi frequency is 
\begin{equation}
    \hbar \Omega = \sum_k \frac{\braket{e|D|k}\braket{k|D|g}}{\hbar \omega_0 - E_k},
\end{equation}
with the summation extending over intermediate states that are connected to the $\ket{g}$ and $\ket{e}$ states. The $E1$ Hamiltonian is 
\begin{align}
    D  = \mathcal{E}_0 \left( \cos{\theta} \, d_0 - \frac{1}{\sqrt{2}} \sin{\theta} e^{i\phi} d_{1} + \frac{1}{\sqrt{2}} \sin{\theta} e^{i\phi} d_{-1} \right),
\end{align}
where $\mathcal{E}_0$ is the laser electric field amplitude, $d_q (q=0,\pm1)$ are the spherical components of the dipole moment operator, and the polarization angles $\theta,\phi$ are defined in the main text.

Using the Wigner-Eckart theorem, matrix elements of the spherical components of the dipole moment can be written as 
 \begin{equation}
     \braket{J_k,m_k|d_q|J_i,m_i} = (-1)^{J_k-m_k}  \left( \begin{array}{ccc} J_k & 1 & J_i \\
 -m_k & q & m_i \end{array}  \right) \braket{J_k||d||J_i}.
 \end{equation}

The state $\ket{g}$ has $J=0,m=0$ and the state $\ket{e}$ has $J=2,m=0$. 

Only three sets of matrix elements between these states and intermediate states result in non-zero contributions to the sum:
\begin{align}
\braket{2,0|d_0|1,0} \braket{1,0|d_0|0,0}  & = \frac{2}{\sqrt{30}}\frac{1}{\sqrt{3}} \, \braket{J_f||d||J_k}\braket{J_k||d||J_i},\\
\braket{2,0|d_1|1,-1} \braket{1,1|d_{-1}|0,0} & = \frac{1}{\sqrt{30}}\frac{1}{\sqrt{3}}e^{i\phi} \braket{J_f||d||J_k}\braket{J_k||d||J_i},\\
\braket{2,0|d_{-1}|1,1} \braket{1,-1|d_1|0,0} & = \frac{1}{\sqrt{30}}\frac{1}{\sqrt{3}}e^{i\phi} \braket{J_f||d||J_k}\braket{J_k||d||J_i}.
\end{align}
All three of these pairs enter the Rabi frequency with the same energy denominator. Therefore 
\begin{align}
    \braket{e|D|k}\braket{k|D|g} &= \mathcal{E}_0^2 \frac{1}{\sqrt{30}}\frac{1}{\sqrt{3}} \braket{J_f||d||J_k}\braket{J_k||d||J_i}  \left[ 2\cos^2{\theta} - 2\left(\frac{1}{\sqrt{2}} \sin{\theta} \, e^{i\phi}\right)^2   \right]\\
    &\propto \braket{J_f||d||J_k}\braket{J_k||d||J_i}  \left( 2\cos^2{\theta} - \sin^2{\theta} \, e^{2i\phi} \right),    
\end{align}
and the two-photon Rabi frequency is
\begin{equation}
    \hbar \Omega \propto  \left( 2 \cos^2{\theta} - \sin^2{\theta}e^{2i\phi} \right) \sum_{k} \frac{\braket{J_f||d||J_k}\braket{J_k||d||J_i}}{\hbar \omega_0 - E_k}.
\end{equation}

For $\phi = 0$ (pure linear polarization), the Rabi frequency $\Omega$ goes to zero when $\cos^2 \theta = \frac{1}{3}$, as shown in Fig.\ \ref{fig:rabi}. For other values of $\phi$, the magnitude of $\Omega$ does not become zero for any value of $\theta$.

\end{document}